\documentstyle[12pt,moriond,psfig]{article}
\begin{document}
\heading{An optical and near-IR survey\\ of nearby clusters of galaxies}

\author{S. Andreon $^{1}$, J. Annis $^{2}$, J.-C. Cuillandre $^{3}$, 
E. Davoust$^{4}$, J. P. Gardner$^{5}$,\\
Y. Mellier$^{6}$, J.-M. Miralles$^{4}$, R. Pell\'o$^{4}$, P. Poulain$^{4}$} 
{$^{1}$ Osservatorio di Capodimonte, Naples, Italy}
{$^{2}$ Fermi Lab., Batavia, Illinois (USA)}
{$^{3}$ CFHT corp., Hawaii, USA}
{$^{4}$ Observatoire Midi Pyr\'en\'ees, Toulouse, France}
{$^{5}$ NASA-GSFC, USA}
{$^{6}$ IAP, Paris, France}
\centerline{\psfig{figure=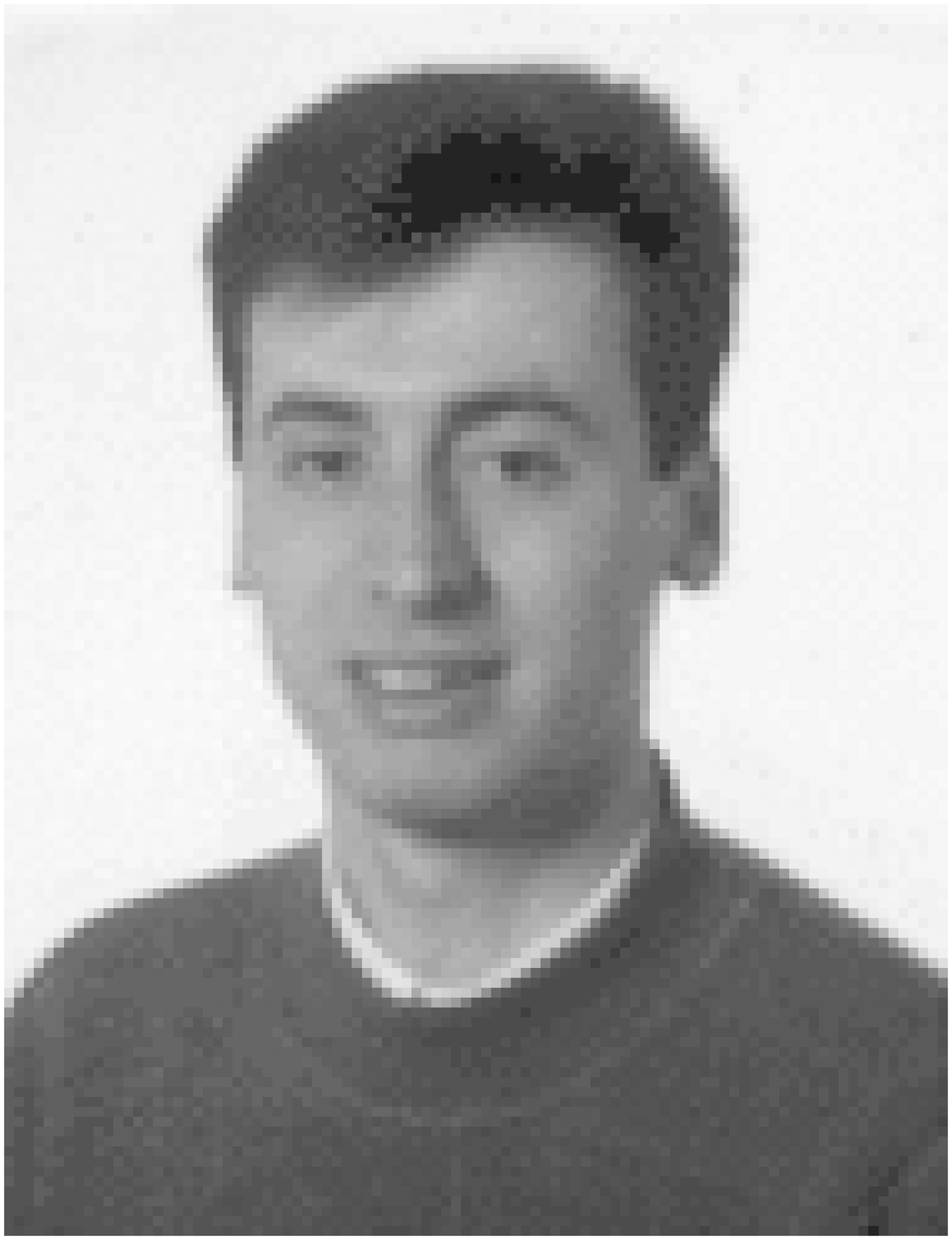,height=3.8truecm}}

\begin{moriondabstract}

We present an optical and near--infrared survey of galaxies in nearby
clusters aimed at determining fundamental quantities of galaxies, such as
multivariate luminosity function and color distribution for each Hubble
type. The main characteristics of our survey are completeness in absolute
magnitude, wide wavelength coverage and faint limiting magnitudes. 

\end{moriondabstract}

\section{Introduction}

Because panoramic detectors have only recently been available in the
near-infrared, our knowledge of infrared based quantities is extremely
incomplete, even for the most fundamental quantities such as the
luminosity function of galaxies (LF) and the color distribution for the
different Hubble types. The cosmological importance of these quantities is
further enhanced because the near-infrared domain is a good tracer of the
mass distribution of galaxies (e.g. Gavazzi et al. 1996) thanks to its low
sensitivity to short time-scale phenomena such as starbursts. Up to a
redshift of $\sim 1$, the K-band samples the spectral energy distribution
beyond $1 \mu m$ which is basically dominated by low-mass stars (Bruzual
\& Charlot 1993). 

The dependence of the near infrared LF on morphological type is still
unknown, especially at faint luminosities, whereas a great effort is
presently being made in the visible (e.g. Lobo et al. 1997). The
comparison between these two LFs for clusters at different redshifts can
provide new constraints on the evolutionary scenarios for galaxies in
clusters, in particular the evolution of the mass distribution as a
function of morphological type, and the relative significance of the
different interactions between cluster galaxies and their environment:
merging processes (Toomre \& Toomre 1972), harassment (Moore et al. 1996),
ram-pressure stripping (Gunn \& Gott 1972), etc.  Furthermore, the nature
of faint blue galaxies seen on deep galaxy counts in the blue band, and of
their possible counterparts in the nearby ($z<0.1$) universe (see the
reviews by Koo \& Kron, 1992, and Ellis 1997) are still an open question.
Finally, the study of the global properties of galaxies in nearby clusters
(color distribution, LF, segregation, etc.) is essential for comparisons
with distant clusters, now observed with the Space Telescope and Keck, in
order to draw an evolutionary picture. 

\setbox1=\vbox{\psfig{figure=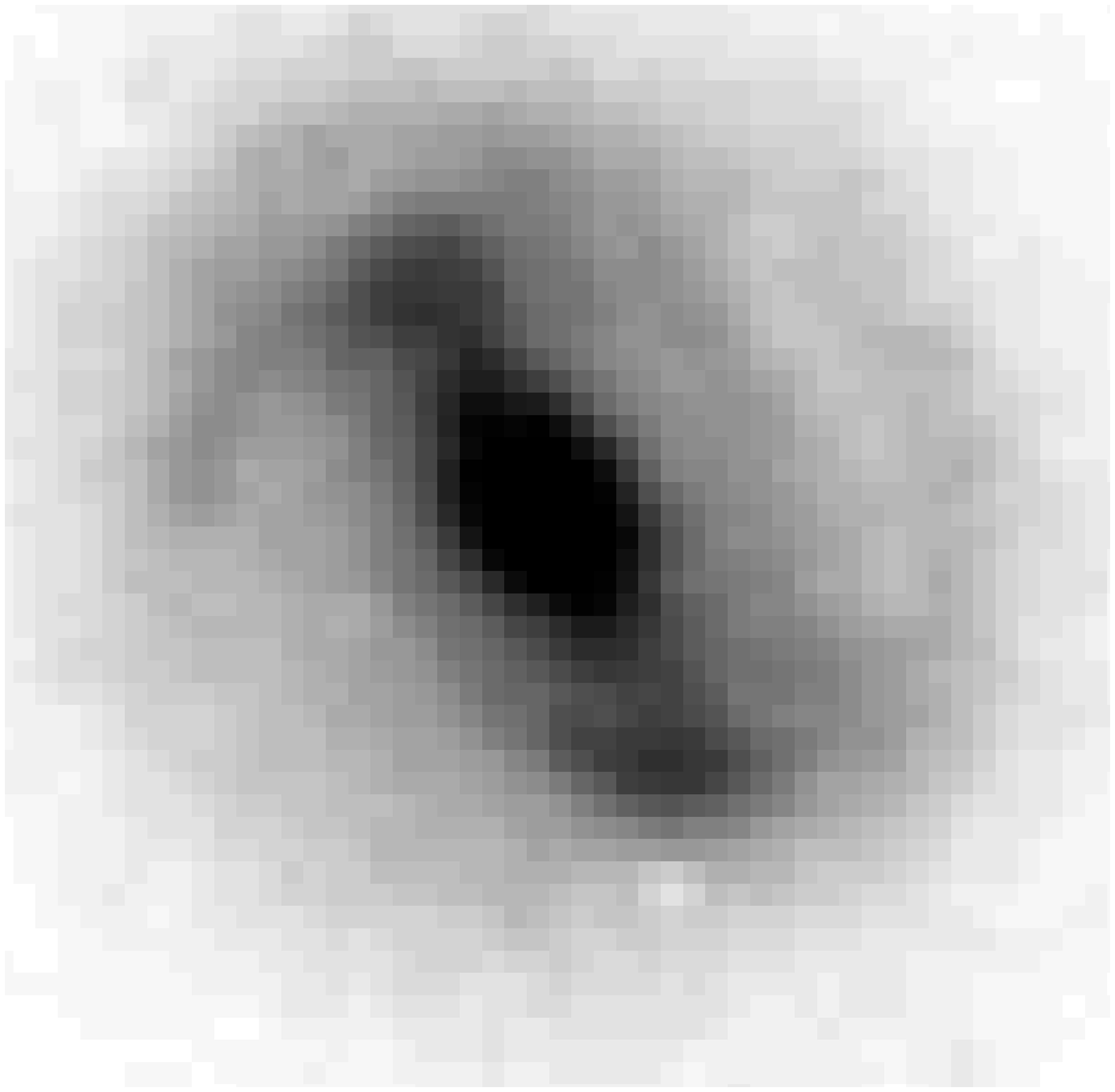,width=8truecm}}
\setbox2=\vbox{\psfig{figure=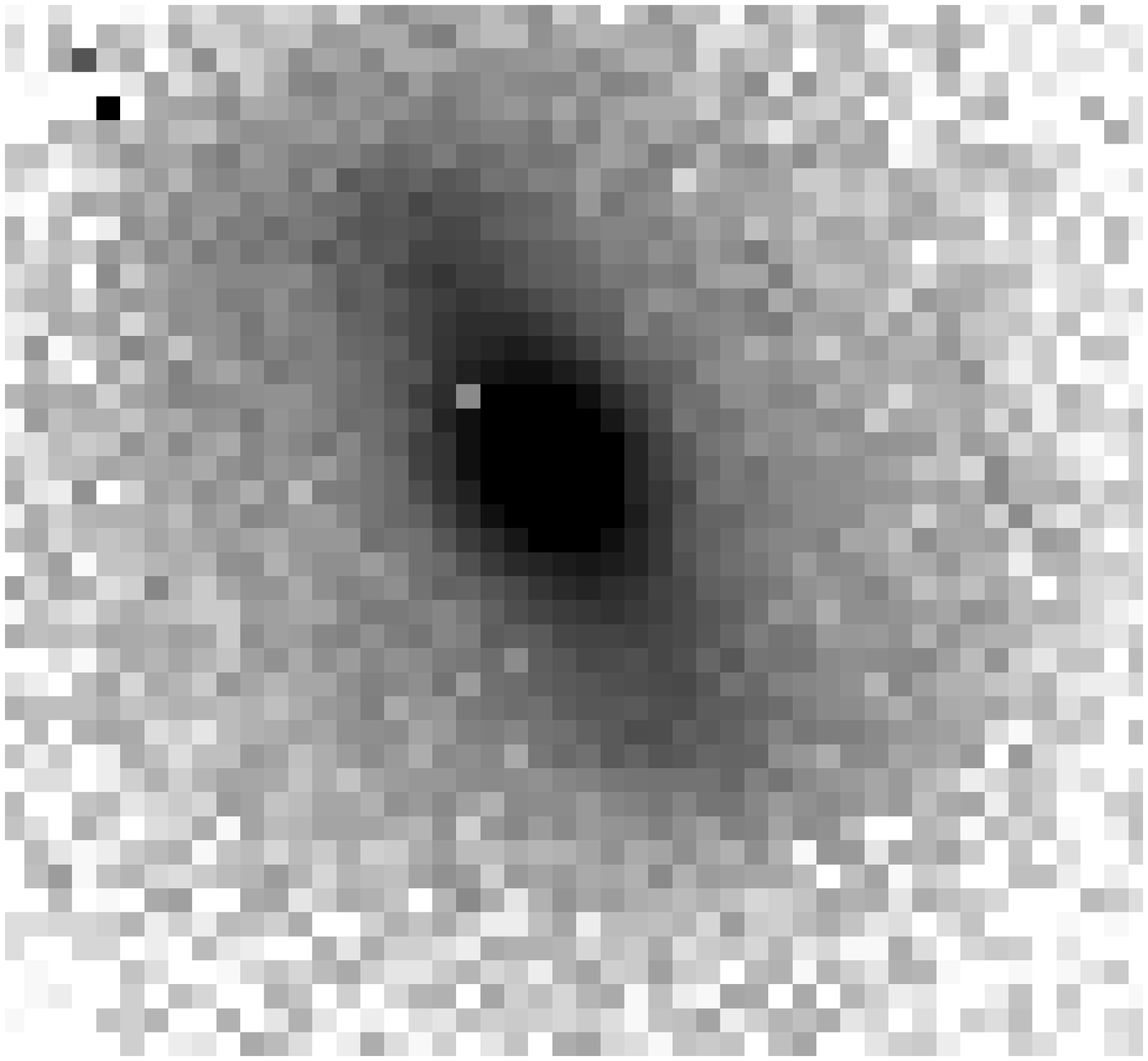,width=8truecm}}

\begin{figure}
\noindent
\centerline{\box1 \box2}

\noindent 
\caption{Comparison between the $B_J$ (left, from POSS-II) and $H$ (right,
Mo\"\i cam data)
morphologies of a galaxy in Coma. The field of view of each panel is 50
arcsec. Sampling used for this figure is 1 arcsec in $B_J$ and in $H$. 
The $H$ image is only preliminarily reduced.} 
\end{figure}

\section{Our survey}

For all these reasons, we started an observational program aimed at
determining basic properties of galaxies in clusters, such as multivariate
luminosity functions, color distributions, type dependence of the
luminosity function, spectrophotometry of the different Hubble types, etc.
The main characteristics of our survey are : 

\begin{itemize}

\item
completeness in absolute magnitude, 

\item
possibility of selecting
 complete samples in various filters, possibly from $B_J$
($\sim  4000$ \AA \ ) to $K$ (2.2 $\mu m$), and

\item
faint limiting magnitudes, in order to reach the typical luminosities of dwarf
galaxies. 
\end{itemize}

We are presently focusing our attention on two clusters: Coma, the
observations of which are still in progress, and Abell 496, for which the
analysis is underway and some preliminary results are available.

\subsection{Coma}

For Coma, we are presently imaging at the 2m telescope at Pic du Midi a
20$\times$20 arcmin region, corresponding to 800$\times$800 Kpc at the
distance of Coma, in $J$, $H$ and $K'$, down to 6 mag below $M^*$ ($B_J
\sim 21$). The region is centered 15 arcmin NE from the cluster center. A
preliminary analysis of the observations in $H$ shows that the data
already acquired reach the expected magnitude limit. The camera used for
this survey is {\it Mo\" \i cam}, set at the F/8 focus, with a pixel of
$0".5$ and a field of $120 \times 120$ arcsec.  The detector is a Nicmos
with good cosmetics and linearity in the relevant intensity range, as well
as a faint level of image residuals ($ \le 3 \% $ between 2 successive
exposures).  The seeing usually ranges from $<$1 to 2 arcsec.  For the
same region, we have obtained CCD images in the optical range at the 2m
telescope of Pic du Midi, as well as prime focus and large-scale
photographic plates, taken with different telescopes (4m at Kitt Peak,
...) and digitized. Figure 1 shows a representative image for a relatively
bright spiral galaxy ($B$=15 mag) in $B_J$ (from POSS-II) and in $H$. 

The Coma data are particularly well suited for the study and the
comparison of galaxy morphologies and photometric properties in different
filter-bands.  In particular, we expect to measure the infrared
multivariate LF down to the luminosity of dwarfs, and to obtain the
color--magnitude diagram in many colors, since we have magnitudes measured
from 2000 \AA \ to 2 $\mu$. The infrared-visual color gradients of
galaxies will also be available from these data, for measuring effective
radii or brightnesses, and the growth curves for different morphological
types. 

\subsection{A496}

For Abell 496, we have obtained a mosaic in $K$, 40$\times$40 arcmin
large, which samples not only the central part of the cluster, but also
the peripheral region (up to 5 cluster core radii), allowing environmental
studies. $K$ images were taken at the 1.5m at CTIO. The images are deep
enough to allow the construction of galaxy samples complete down to 5 mag
below $M^*$. The near-infrared luminosity function of this cluster is
presently available (Gardner \& Annis 1997, in preparation). 

In the optical, we have at our disposal three deep overlapping pointings
of 14$\times$14 arcmin taken with Mocam (Cuillandre et al. 1996) at the
CFHT in the $V$ filter. The images are very deep, and are the basis of our
object catalog ($\sim 90\%$) complete down to $V=25$ ($\sim M^*+9$). The
images were taken under good seeing conditions (0.7-0.8 arcsec FWHM) which
allows an easy star--galaxy discrimination down to $V=24$ mag. At fainter
magnitudes, the contribution of stars to the sample is negligible.  These
images were taken through thin cirrus. In order to put the CFHT images on
an absolute photometric scale, we re-observed several selected galaxies in
this field at the 2m Pic du Midi telescope.  The S/N ratio of these CFHT
images is so good (the detection threshold at $2 \sigma$ is $\mu_V=25$ mag
arcsec$^{-2}$) that it allows one to easily detect the large majority of
low-surface brightness galaxies of the same type as those already known
in Virgo (Impey, Bothun \& Malin 1988) and elsewhere (Dalcanton et al. 
1997). 

A deep image of Abell 496 in the $I$ filter was also taken with the UH8K
camera (Cuillandre et al. 1996) at the CFHT under similar conditions. The
field of view of the image is 28$\times$28 arcmin. A CCD Schmidt frame was
taken at the Calern observatory in order to photometrically calibrate this
mosaic image. 

\setbox1=\vbox{\psfig{figure=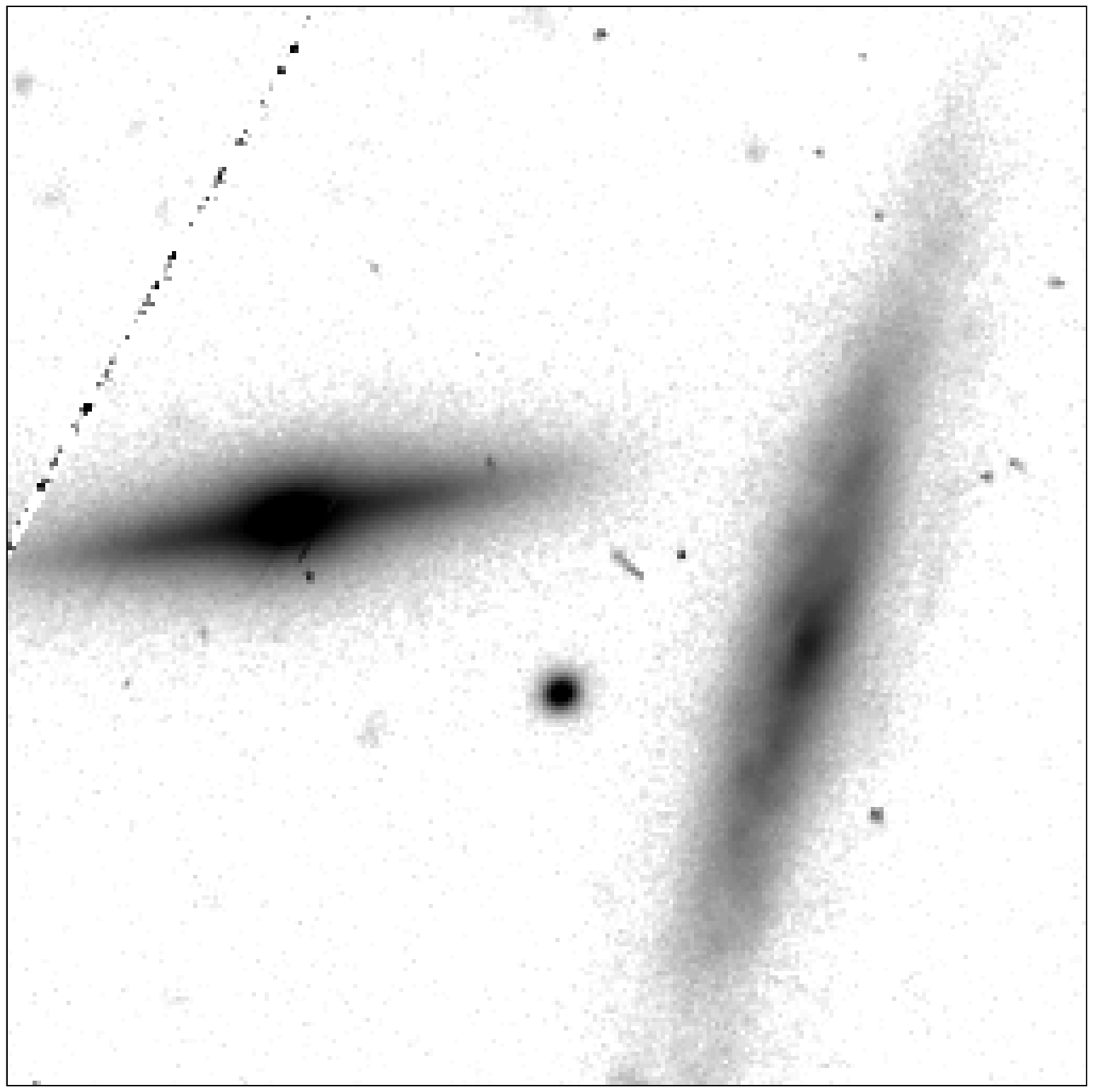,width=8truecm}}
\setbox2=\vbox{\psfig{figure=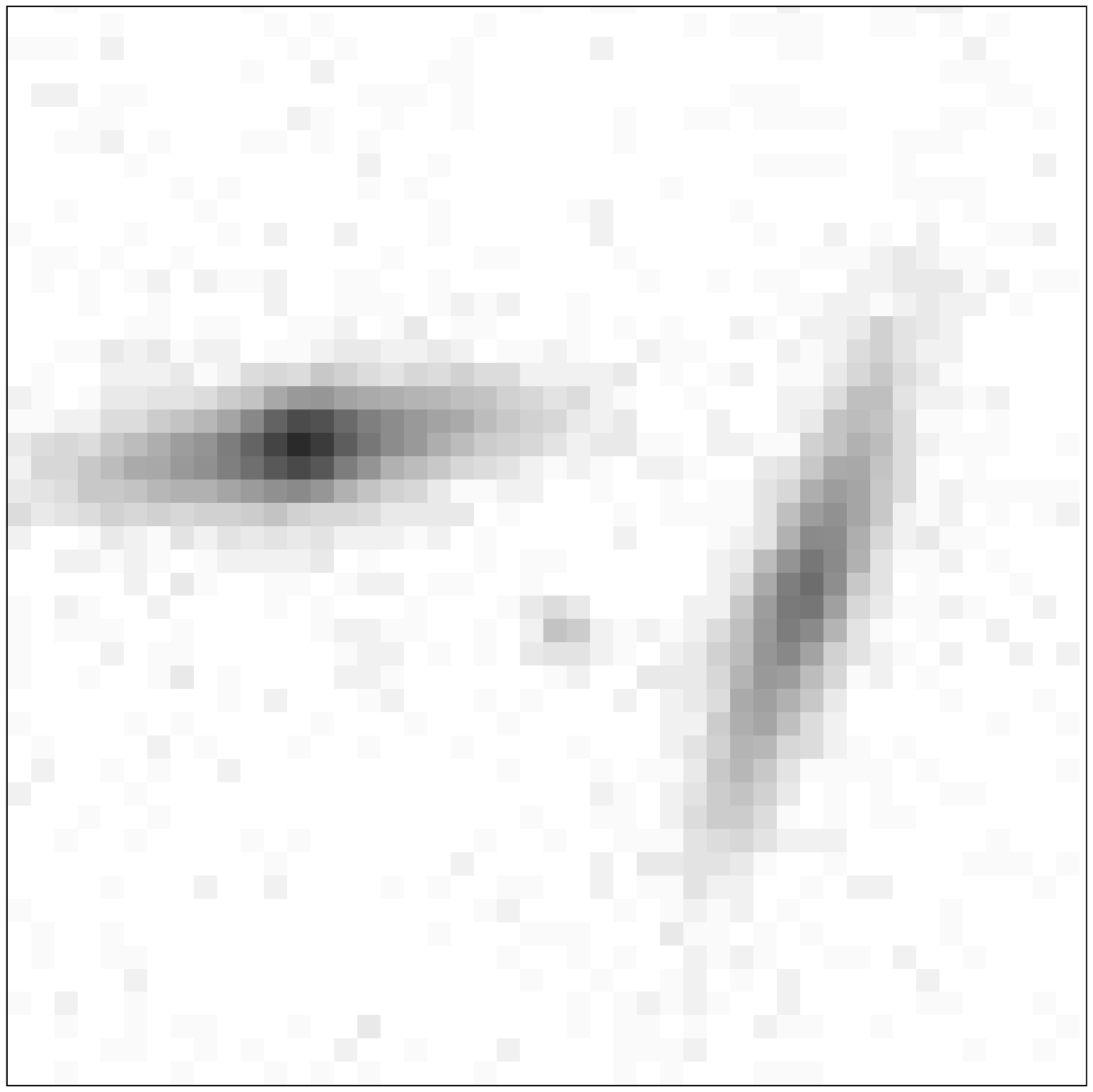,width=8truecm}}
\setbox3=\vbox{\psfig{figure=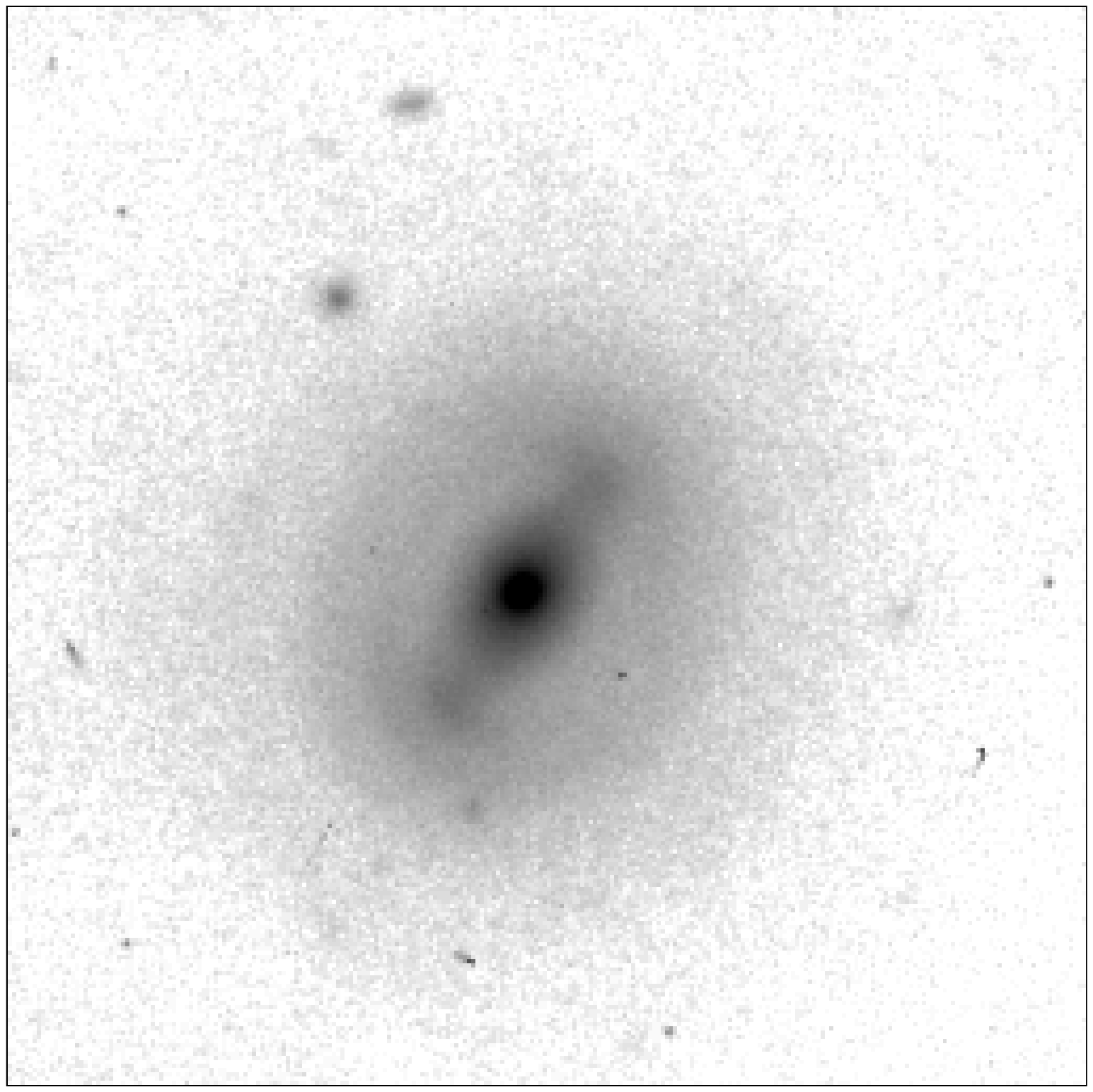,width=8truecm}}
\setbox4=\vbox{\psfig{figure=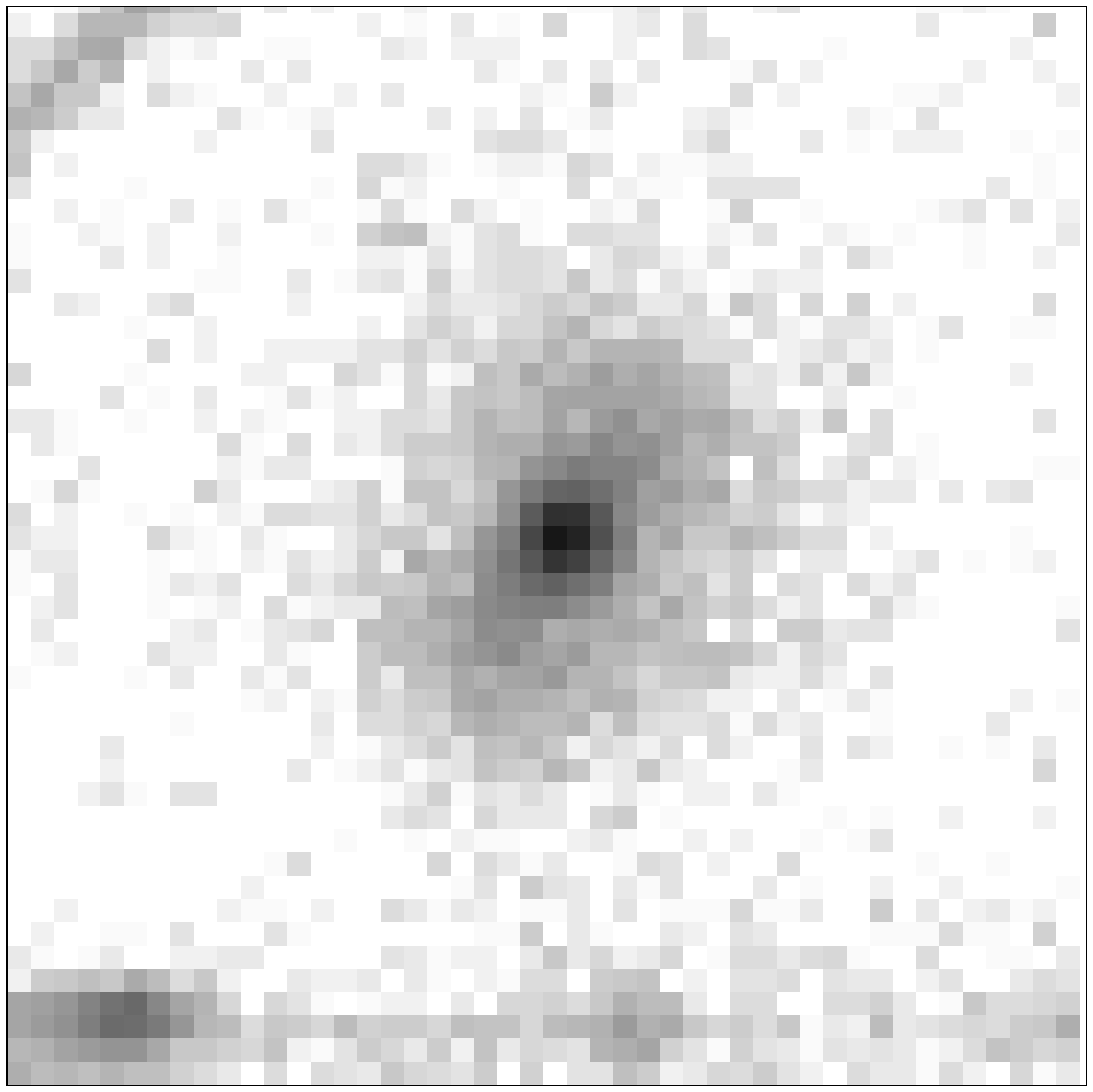,width=8truecm}}

\begin{figure}
\noindent
\centerline{\box1 \box2}
\noindent
\vskip -2 truecm
\centerline{\box3 \box4}

\vskip -2 truecm\noindent \caption{Comparison between the $V$ (left) and $K$
(right) morphologies of galaxies in Abell 496. The field of view of each
panel is 50 arcsec. The seeing is 0.7 and 1.7 arcsec for $V$ and $K$
respectively. The sampling is 0.2 and 1.1 arcsec for $V$ and $K$ .}
\end{figure}

\section{Preliminary results}

\subsection{Galaxy morphology}

Figure 2 shows the morphology of some typical bright galaxies of Abell
496. In general, the visual appearance in $K$ is quite similar to that of
cluster galaxies in the optical bands : bars, bulges, arms, rings and
halos are present and they are in no way peculiar. The direct comparison
of the morphologies in the $K$ and $V$ filters for the small sample of
large galaxies in the central part of Abell 496 shows that all the
morphological details present in $V$ are also found in the near infrared
and vice versa. In particular, none of these galaxies appears of earlier
type in the infrared than in the optical. However, the sampling and seeing
are much poorer in $K$ than in $V$, which makes the comparison difficult,
and the majority of the sample galaxies are Es or S0s, i.e. galaxies
dominated by an old stellar population. 

This comparison between optical and near-IR morphologies is much easier
for the galaxies in Coma, as shown in the example of Figure 1. A more
quantitative comparison of the morphological appareance of galaxies as a
function of wavelength is presently under way. 

\begin{figure} 
\psfig{figure=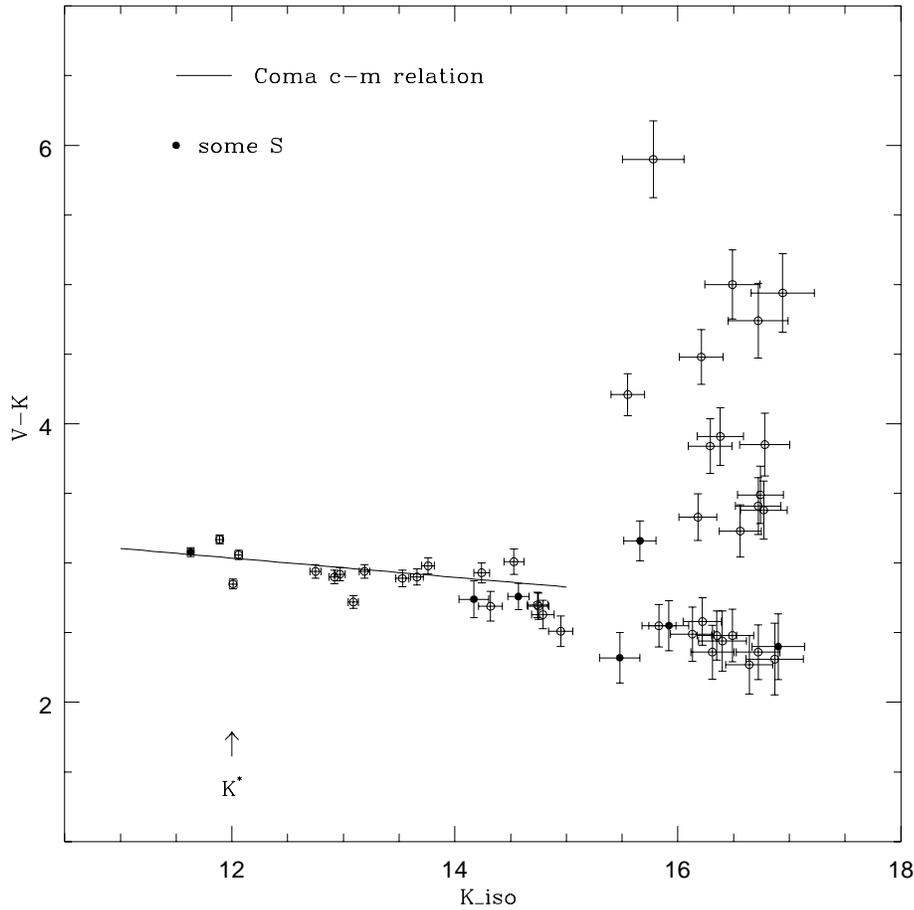,width=13.5truecm} 
\caption{
Color--magnitude diagram for the galaxies in the direction of the Abell 496
cluster. The sample is selected and complete in $K$. Filled dots are
morphologically identified spiral galaxies. The plotted line is the 
color-magnitude relation in Coma from Bower, Lucey \& Ellis (1992), 
corrected for the difference in distance modulus between 
Coma and Abell 496.} 
\end{figure}

\subsection{Color--magnitude diagram}

Figure 3 shows a preliminary version of the color-magnitude diagram of
galaxies in the central part of Abell 496. The sample is selected and
complete in the $K$ filter. Colors are measured within a 4 arcsec aperture
(3.7 kpc for galaxies in the Abell 496 cluster), which only includes the
central part of the brightest galaxies but all the $S/N >1$ region of
faint galaxies. At the brightest magnitudes, the color magnitude relation
of Abell 496 is in excellent agreement with that of Coma (Bower, Lucey \&
Ellis 1992), which is plotted in the figure as a line, after correction
for the difference in distance modulus between the two clusters. The
scatter is small, 0.2 mag, and not much larger than the photometric
errors. At faint magnitudes, there is a population of red (say $3 \lsim
V-K\lsim 6$) galaxies.  Such extremely red galaxies are also found in
multicolor field surveys (e.g. Moustakas et al. 1997), but at fainter
magnitudes (K=19-20), and they are identified in this case to early-type
galaxies having undergone a burst of star formation at z$>$5 and viewed at
z$\sim$1. Their total number in the field of Abell 496 is of the same
order as that expected for the background (according to the field counts
by Gardner et al. 1996). Nevertheless, we cannot exclude that at least
some of these galaxies belong to the cluster. A straightforward
calculation shows that a dusty Sb/Sc galaxy, with $A_V \sim 1-2$, will
have $V-K \sim 3.7$ to $5$. While waiting for a spectroscopic
confirmation, we can discriminate between the two hypotheses by studying
the clustercentric radial dependence of the density of these galaxies.

\acknowledgements{SA thanks the organizers of the "Rencontres des Moriond"
for financial support to attend the congress.}

\end{document}